\documentclass{optica-article}

\journal{opticajournal} % for journals or Optica Open

\articletype{Research Article}

\usepackage{amssymb}
\usepackage{lineno}
\usepackage{changes}
\usepackage{amsmath,mathtools}

\usepackage{soul}

\begin{document}

\title{Hearing carrier-envelope offset frequency and phase in air with a microphone}

\author{Meng Han,\authormark{1,*} Ming-Chang Chen,\authormark{2,*} Ming-Shian Tsai,\authormark{2} and Hao Liang \authormark{3}}

\address{\authormark{1}J. R. Macdonald Laboratory, Department of Physics, Kansas State University, Manhattan, Kansas 66506, USA\\
\authormark{2}Institute of Photonics Technologies, National Tsing Hua University, Hsinchu 300044, Taiwan\\
\authormark{3}Max Planck Institute for Physics of Complex Systems, Nöthnitzer Straße 38, 01187, Germany}

\email{\authormark{*}meng9@ksu.edu; mingchang@mx.nthu.edu.tw} %% email address is required; see note below about the corresponding author designation

% use {asbstract*} to suppress the copyright line. Copyright information will be added in production

\begin{abstract*} 
Attosecond science and frequency metrology rely on the precise measurement and control of the laser pulse waveform, a feat traditionally achieved using optoelectronic techniques. In this study, we conducted a laser-induced acoustic experiment in air ionized by carrier-envelope phase (CEP)-stabilized sub-4 femtosecond pulses. Our results reveal that the acoustic signal exhibits CEP dependence in few-cycle pulses, primarily through amplitude modulation from laser-driven ionization. This novel optoacoustic phenomenon enables not only the measurement of the carrier-envelope offset frequency but also the direct characterization of the waveform of optical pulses through a microphone. Our study highlights the potential of laser-induced acoustic waves for advancing frequency metrology and ultrafast science.
\end{abstract*}

%%%%%%%%%%%%%%%%%%%%%%%%%%  body  %%%%%%%%%%%%%%%%%%%%%%%%%%
\section{Introduction}
%The harmonic oscillator is one of the most fundamental physical models, extensively used to describe laser-driven bound and free electronic systems. When the light intensity is sufficiently strong to introduce some distortions on the simple harmonic motion, nonlinear interactions can generate new frequencies beyond the spectrum of the driving field, significantly making our world "colorful" with nonlinear phenomena, such as sum frequency generation \cite{franken1961generation}. Laser-induced plasma is a typical extremely nonlinear system, which can generate high-order harmonics in the extreme-ultraviolet \cite{ferray1988multiple} and soft-x-ray \cite{spielmann1997generation,chang1997generation} regimes, as well as terahertz radiation through four-wave optical rectification \cite{cook2000intense,kress2004terahertz,xie2006coherent} or transient photocurrent \cite{kim2007terahertz,gildenburg2007optical}. These new frequencies not only provide spectroscopic tools for studying the mechanisms underlying their generation but also serve as novel light sources for probing various dynamic processes in other systems \cite{corkum2007attosecond,jordan2020attosecond}.

The carrier-envelope phase, CEP, is a critical parameter for controlling the electric-field waveform of optical light pulses. For few-cycle femtosecond pulses, the CEP can modulate the electric-field waveform within the intensity envelope, transitioning it from a "sine" shape to a "cosine" shape. This capability enables the generation of single isolated attosecond light pulses \cite{hentschel2001attosecond,chini2014generation}. Accurate measurement of CEP primarily relies on the f-to-2f technique \cite{jones2000carrier,apolonski2000controlling,kakehata2001single,baltuska2003phase} and its variants \cite{fuji2005monolithic,yu2016monolithic}. Alternative methods include the stereo photoelectron phase meter \cite{paulus2001absolute}, which derives CEP information by analyzing spatial asymmetry in electron emissions during above-threshold ionization using two detectors. However, this approach requires complex apparatus and high-vacuum conditions. Another advanced technique, nanotip ionization \cite{putnam2017optical,keathley2019vanishing}, utilizes the enhanced local electric field at a nanotip to induce tunneling ionization, detecting the CEP via the resulting current. Recently, photoconductive current measurements in air have been employed as a simpler alternative for CEP detection in circular polarization \cite{kubullek2020single}. Regarding the carrier-envelope offset frequency ($f_{\rm{ceo}}$) \cite{ye2005femtosecond}, which represents the rate of change of the CEP, it is predominantly measured using the f-to-2f technique.

Laser-induced acoustic wave generation has been studied since the 1960s \cite{nelson1964experimental}, shortly after the invention of the laser. Previous studies \cite{kaleris2024laser,kaleris2024laser1,kaleris2021laser,lengert2022optoacoustic,kaleris2020correlation,kaleris2019experimental,manikanta2016effect,hosoya2013acoustic,oksanen1994photoacoustic} typically utilized pulse durations ranging from several tens of femtoseconds to picoseconds and nanoseconds. In these cases, the CEP of light pulses was not stabilized and hence the sensitivity of acoustic wave to the optical waveform of driving pulses is not completely revealed. In this work, we present the first laser-acoustic experiment using CEP-stabilized sub-4-fs light pulses \cite{balvciunas2011carrier,tsai2022nonlinear}. Our experiments revealed that the waveform of laser-induced acoustic waves depends on the CEP of the driving light pulses, primarily affecting their amplitude. We primarily attribute the CEP dependence to the modulation of the ionized hot electron density, which influences the gas temperature and pressure after strong-field ionization. This new optoacoustic phenomenon provides an alternative for directly sensing $f_\text{ceo}$ using a microphone. By leveraging this phenomenon, we also introduce a pulse characterization method that effectively "hears" light in ambient air. 

\section{Results}

\begin{figure}[htbp]
\centering
\includegraphics[width=\linewidth]{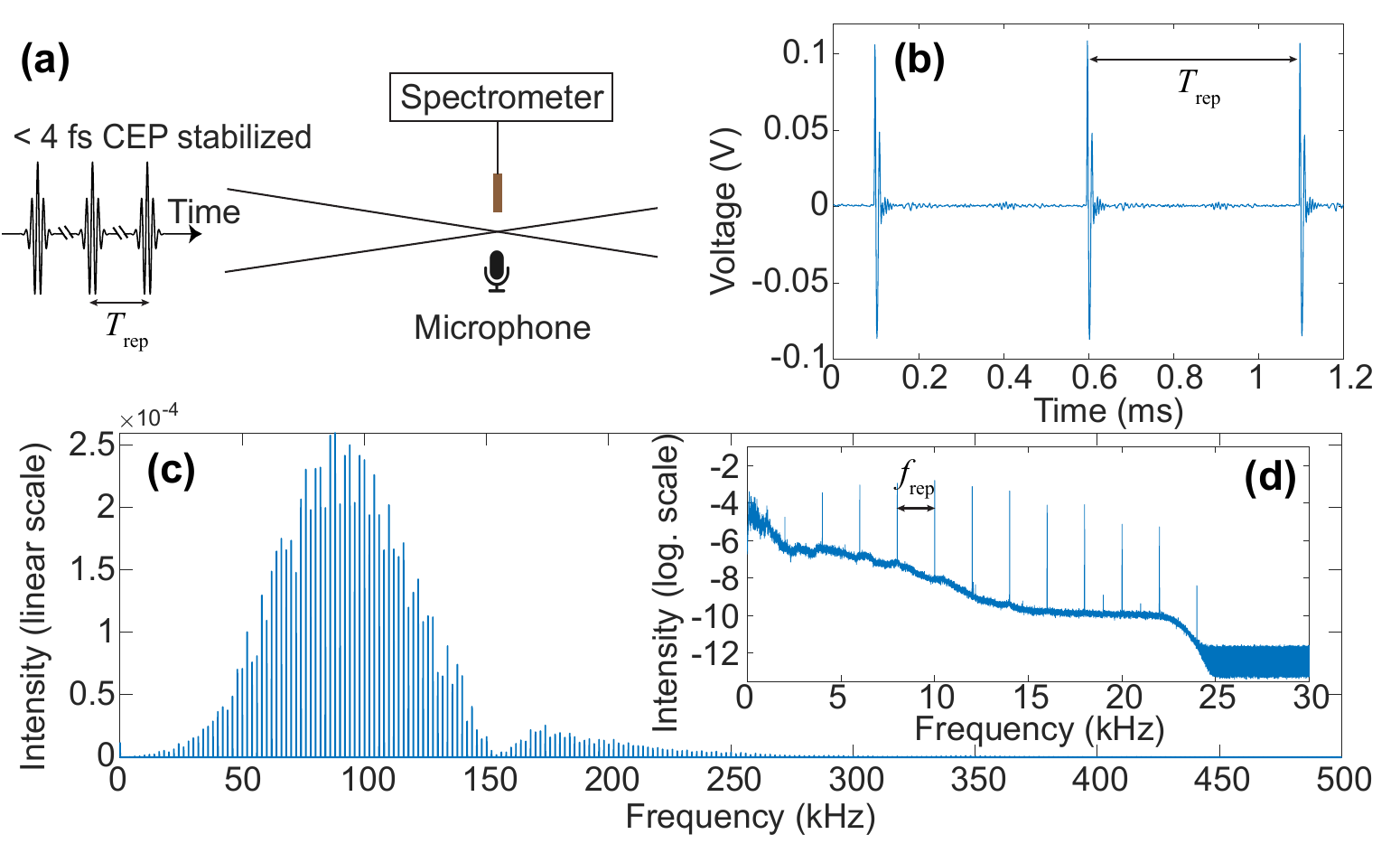}
\caption{\textbf{Laser-induced acoustic wave generation} (\textbf{a}) Experimental setup. The acoustic waves and side-emission fluorescence are measured simultaneously as a function of CEP. $T_{\textrm{rep}}$ denotes the repetition period of the optical pulses. textbf{b} Measured acoustic waveform in the time domain from a state-of-the-art microphone. textbf{c}) Spectral intensity of the acoustic waves, obtained by applying the Fourier transform to the waveform in (\textbf{b}).  (\textbf{d}) Measured spectral intensity of acoustic waves by a common housekeeping microphone. $f_{\textrm{rep}}$ represents the repetition rate.} 
\label{fig:figure1}
\end{figure}

As shown in Figure 1(a), we focused CEP-stabilized few-cycle pulses in air, with a center wavelength of \(\sim 900\) nm, operating at \( f_{\text{rep}} = 2.0008 \) kHz (for details, see the Supplementary document and \cite{tsai2022nonlinear}). The peak intensity at the laser focus exceeds $1\times10^{14}$ W/cm$^2$, resulting in a plasma filament. A microphone and a spectrometer were positioned on opposite sides of the filament to capture the laser-induced sound wave and the side-emission fluorescence, respectively. The acoustic wave is considered to originate from the gas pressure fluctuation due to the heat relaxation after strong-field ionization \cite{kaleris2021laser}. Fluorescence is an another important mechanism for the excited molecules releasing the energy in the laser-induced plasma. Both signals should be proportional to the total ionization probability. Hence, acoustic waves and side-emission fluorescence are measured simultaneously as a function of laser CEP (Figure 2). Figure 1(b) presents the measured voltage signal from a state-of-the-art microphone with a dynamic range of up to 170 dB as a function of time, representing the evolution of gas pressure. The acoustic waveform resembles an N-shaped pulse \cite{oksanen1994photoacoustic}, characteristic of a blast wave. The time interval between two consecutive acoustic blasts corresponds to the spacing of the driving pulses, $T_{\rm{rep}} = 1/f_{\rm{rep}}$. This periodicity gives rise to a frequency comb in the spectrum of the acoustic wave. In Figure 1(c), we show the spectral intensity of the acoustic comb. The frequency spacing of the acoustic comb teeth corresponds to the laser repetition rate and the order can exceed one hundred. Notably, these acoustic spectra are partly audible to the human ear. Figure 1(d) shows the measured spectral intensity by a common housekeeping microphone, where the cutoff frequency is around 22 kHz, aligning with the frequency range of human hearing. All optoacoustic phenomena reported here are highly robust and can be observed using a common housekeeping microphone. More experimental details are presented in the Method section of Supplemental document.

\begin{figure}[htbp]
\centering
\includegraphics[width=\linewidth]{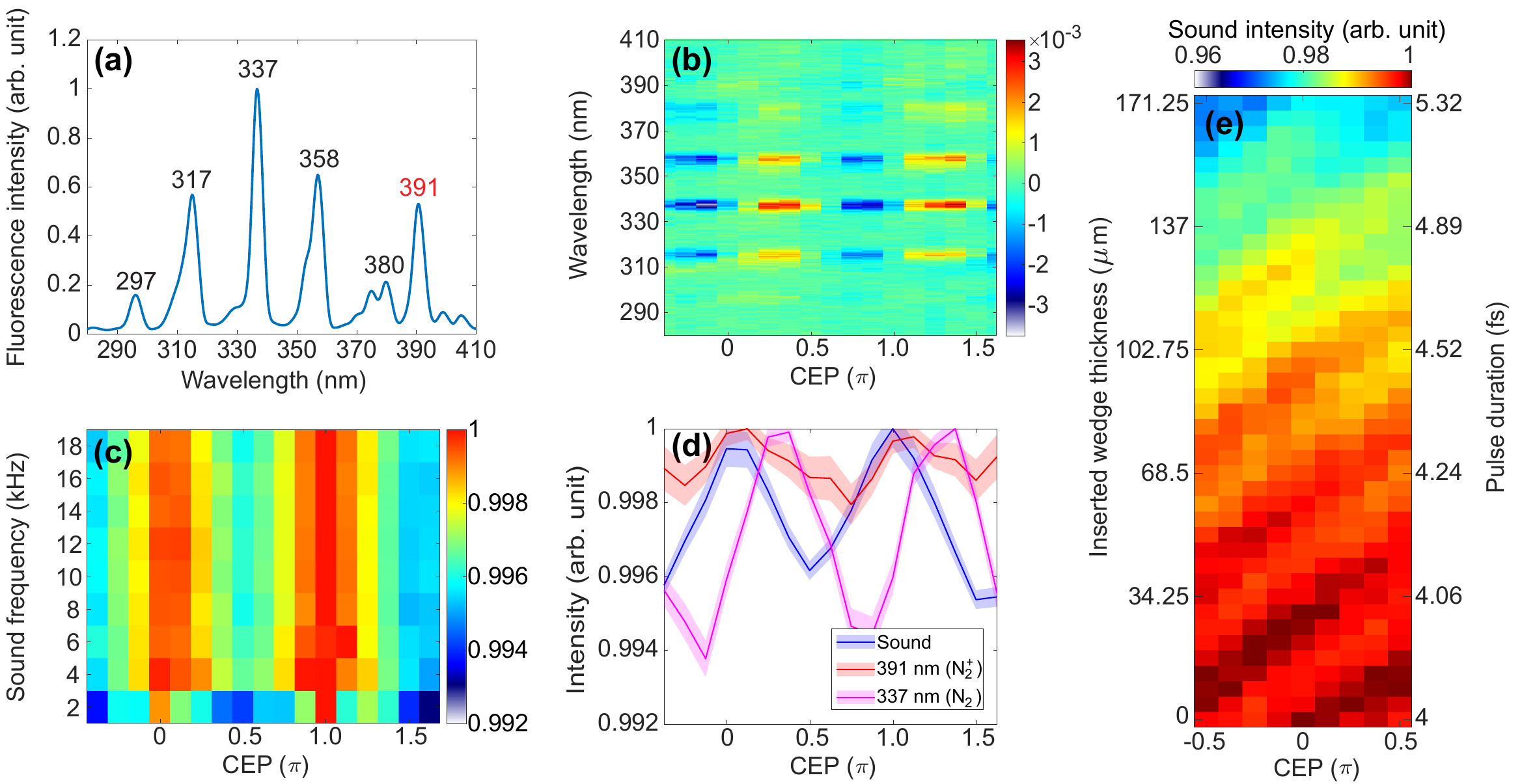}
\caption{\textbf{CEP-dependent fluorescence and acoustic wave generation in air.} (\textbf{a}) CEP-averaged fluorescence spectrum. (\textbf{b}) CEP-resolved differential distribution of the fluorescence spectrum. (\textbf{c}) Acoustic wave spectra as a function of laser CEP. Each CEP-dependent spectral component is normalized to its maximum value. (\textbf{d}) Comparison of CEP dependence of the total acoustic intensity and the fluorescences from ionized and excited nitrogen molecules. The shaded areas represent the standard deviations of four measurements,indicating that the CEP measurement uncertainty between the f-to-2f and acoustic methods is approximately 80 as, primarily attributed to the CEP uncertainty of our laser ($\sim$100 mrad, 60 as). (\textbf{e}) Two-dimensional CEP scan of the total sound intensity by varying both the pulse CEP from the laser oscillator and the inserted thickness of a fused silica wedge at the user end station.}
\label{fig:figure2}
\end{figure}

Figure 2(a) shows the measured static fluorescence spectrum in the background-free ultraviolet regime, where the photon lines can be assigned into two categories. The dominant category is the transition from the C$^3\Pi_u$ state to the B$^3\Pi_g$ state in excited neutral $N_2$ molecules, such as 337 nm ($\nu =0 \rightarrow \nu'= 0$), 358 nm ($\nu =0 \rightarrow \nu'= 1$), 317 nm ($\nu =1 \rightarrow \nu'= 0$), 297 nm ($\nu =2 \rightarrow \nu'= 0$) and 380 nm ($\nu =0 \rightarrow \nu'= 2$). The second category is the transition from the B state to the X state in $N_2^+$ ions, such as 391 nm ($\nu =0 \rightarrow \nu'= 0$). $\nu$ and $\nu'$ donate the vibrational quantum numbers of the upper and lower states, respectively. Figure 2(b) displays the CEP-resolved differential distribution of the fluorescence spectrum, which is defined as [$Y(\phi_{\rm{CEP}},\lambda)$ -  $\overline{Y}(\lambda)$]/$\overline{Y}(\lambda)$, where $Y(\phi_{\rm{CEP}},\lambda)$ and $\overline{Y}(\lambda)$ are the CEP-resolved and -averaged fluorescence spectrum intensities at wavelength $\lambda$. The differential distribution reveals that there is a notable phase shift between the photon lines from $N_2$ and $N_2^+$, indicating their generation mechanisms are different. The fluorescence of neutral $N_2$ relies on collision excitation \cite{xu2009mechanism,mitryukovskiy2015plasma}, while the fluorescence of $N_2^+$ is proportional to the total ionization probability. As expected, the fluorescence of $N_2^+$ aligns with the laser-induced acoustic intensity.
 Velocity mapping imaging (VMI) analysis of gas ionization yield and photoelectron momentum driven by sub-4-fs pulses confirms that a "cosine-shaped" pulse (CEP = 0 or \(\pi\)) produces higher ionization than a "sine-shaped" pulse (CEP = \(\pi/2\) or \(3\pi/2\)). Further VMI results will be presented in a forthcoming publication.  While we know that maximum acoustic wave (ionization) occurs at \(\text{CEP} = 0\) or \(\pi\), it remains unclear whether the absolute value corresponds to \(0\) or \(\pi\). The CEP values in the figures represent relative CEP. 

Figure 2(c) shows the acoustic wave intensities as a function of the laser relative CEP. The first order acoustic spectrum at $f_{\rm{rep}}$ was partly contaminated by the background noise in the lab, therefore the CEP dependence is not perfectly smooth. Starting from the second-order acoustic spectrum ($\approx$ 4 kHz), the CEP dependence closely resembles a standard sine or cosine function and we didn't observe any significant chirps among these spectral components, which demonstrates the teeth in the acoustic frequency comb are phase locked. Figure 2(d) shows a comparison of the CEP dependence between the frequency-integrated sound signal and the 391-nm and 337-nm fluorescence signals. Although the contrast of the CEP dependence is only $\approx$ 0.2\%, it's still detectable with a housekeeping microphone. Our experimental result reveals that the sound signal is nearly synchronized with the ionic signal, against with the signal from excited neutral molecules. This correlation demonstrates that the intensity of the acoustic wave is proportional to the total ionization probability, or equivalently, the electron density in the plasma filament. In Figure 2(e), we display a two-dimensional scan of the total sound intensity as a function of the laser CEP and the thickness of a fused-silica wedge pair inserted into the beam path. The slight difference between the group and phase velocities in fused silica allows for CEP modulation by adjusting the wedge thickness. The diagonal fringes observed in Figure 2(e) indicate an equal and periodic dependence on both the laser CEP and the wedge thickness, confirming the reported CEP dependence of the laser-acoustic wave. Additionally, by varying the wedge thickness, we introduced dispersion and correspondingly stretched the pulse duration. Our results show that for pulses longer than 5 fs, CEP dependence weakens but remains observable. Here, we would like to emphasize that, unlike conventional CEP measurement techniques such as the f-to-2f interferometer \cite{jones2000carrier,apolonski2000controlling,kakehata2001single,baltuska2003phase}, the stereo photoelectron phase meter \cite{paulus2001absolute}, and nanotip ionization \cite{putnam2017optical,keathley2019vanishing}, which exhibit a
$2\pi$ rad periodicity, the total acoustic intensity in Figure 2(d) shows a $\pi$ rad periodicity with respect to the CEP. This is consistent with the CEP-dependent high harmonic yield in isolated attosecond pulse generation using few-cycle pulses, where the CEP modulation period is also 
$\pi$ rad \cite{baltuvska2003attosecond,tsai2022nonlinear}. This correlation arises because there are two electric field peaks within one optical cycle. Note that this ambiguity of $\pi$ can be removed by introducing a second harmonic to break the symmetry of the light field \cite{mashiko2008double,mashiko2009extreme,park2018direct}.

It is important to note that our driving laser is not an optical frequency comb \cite{canella2024low,huang2011high}, as $f_{\rm{rep}}$ is not actively stabilized. A 15 mHz deviation in the repetition rate was observed, corresponding to a 2 GHz linewidth at $\approx$ 290 THz (1030 nm), which is significantly broader than the 2 kHz comb mode spacing. Nevertheless, in the spectrum of the generated acoustic wave, the observed comb modes indicate that the acoustic pulse-to-pulse time jitter is smaller than the width of individual acoustic teeth. Regarding the synchronization between optical and acoustic pulses, we use the signal from a fast photodiode as a trigger to record the acoustic waveform in the time domain using a 2.5 GHz oscilloscope. The stabilized and repeatable acoustic waveform exhibits a time jitter smaller than the 2.5 GHz sampling interval, indicating that the timing jitter between optical and acoustic pulses is less than 400 picoseconds. However, precisely quantifying the timing jitter—whether in the picosecond, femtosecond, or attosecond range—remains an important topic for future investigation.

To gain insight into the acoustic waveform as a function of CEP, Figure 3(a) shows recorded acoustic waveforms $P_N(\phi_{\textrm{CEP}},\it{t})$, while Figure 3(b) presents the CEP-averaged acoustic waveform $\overline{P_N}(t)$. While it is difficult to discern any clear structures directly from $P_N(\phi_{\textrm{CEP}},\it{t})$, Figure 3(c) displays the differential acoustic waveform defined as [$P_N(\phi_{\textrm{CEP}},\it{t})$ -  $\overline{P_N}(\it{t})$], which highlights the influence of CEP on the acoustic waveform. A stepwise $\pi$ phase shift in the differential acoustic waveform is observed when varying the CEP of the driving optical pulses, as indicated by the dashed curves in Figure 3(c). Notably, while the overall shape of the acoustic waveform remains largely unchanged (Figures 3(b)), the CEP-dependent component (i.e., the differential acoustic waveform) accounts for only $\approx 0.1\%$ of the total energy. Figure 3(d) further highlights the differential waveform lineouts at relative CEP values of $0$ and $0.5\pi$, showing that the differential acoustic waveform closely resembles the overall acoustic waveform (Figure 3(b)) while exhibiting opposite amplitude modulations (a period of $\pi$). This $\approx 0.1\%$ change in the differential waveform explains the $\approx 0.2\%$ CEP-dependent amplitude variation observed in the laser-induced acoustic wave in Figure 2(d). In short, when the CEP of the optical pulse is continuously varied, the observed differential acoustic waveform exhibits a stepwise $\pi$ phase shift rather than a continuous change. This indicates that CEP primarily affects the amplitude of the acoustic wave rather than its phase.
%In addition to Figure 2(d), which demonstrates the $\approx$ 0.2\% CEP-dependent amplitude of the laser-induced acoustic wave, a phase shift in the differential acoustic waveform was also observed when varying the CEP of the driving optical pulses, as indicated by the dashed curves in Figure 3(c). Note that the overall shape of the acoustic waveform remains largely unchanged (see Figure 3(a) and (b)), while the CEP-dependent component (the differential acoustic waveform) accounts for only $\approx$ 0.1\% of the total energy. Figure 3(d) compares the lineouts of the differential waveforms at relative CEP values of 0 and  0.5$\pi$, showing opposite amplitude modulations (a $\pi$ phase shift) in the differential acoustic waveforms. Notably, a  0.5$\pi$ shift in CEP exactly transforms the optical waveform from a "sine" shape to a "cosine" shape, or vice versa. However, the underlying mechanism responsible for this CEP-dependent phase shift in the differential acoustic waveforms remains unclear.}

In the frequency metrology community, the emphasis is placed on $f_\text{ceo}$, as it directly determines the absolute frequency offset of the comb teeth. Here, we demonstrate how to extract $f_{\text{CEO}}$ from CEP-induced variations in acoustic amplitude. The acoustic waveform $P_N$ after the CEP modulation can be expressed as 
\begin{equation}
\begin{split}
    P_N(\phi_{\textrm{CEP}},\textit{t}) = P_0(\textit{t})[1+\sigma \textrm{cos}(2\phi_{\textrm{CEP}})]*\sum_n \delta(t-nT_{\textrm{rep}}) \\
    =P_0(\textit{t})[1+\sigma \textrm{cos}(2\times2\pi\times f_{\textrm{ceo}}t)]* \sum_n \delta(t-nT_{\textrm{rep}}),
\end{split}
\end{equation}
where $P_0(\textit{t})$ is the unmodulated waveform, $\sigma$ is the contrast of the CEP dependence, the sign of * represents the convolution operator, $\delta(t-nT_{\textrm{rep}})$ is the Delta function, and $\phi_{\textrm{CEP}}=2\pi\times f_{\textrm{ceo}}t$. As previously mentioned, the total acoustic intensity in Figure 2(d) exhibits a $\pi$ rad periodicity with respect to the CEP, explaining why $\phi_{\text{CEP}}$ appears with a prefactor of 2.

In the frequency domain, the acoustic spectrum becomes 
\begin{equation}
\begin{split}
    \mathcal{F}(P_N(\phi_{\textrm{CEP}},\textit{t})) = \mathcal{F}(P_0(\textit{t})[1+\sigma \textrm{cos}(2\times2\pi\times f_{\textrm{ceo}}t)])\cdot \mathcal{F}(\sum_n \delta(t-nT_{\textrm{rep}})) \\
    = P_0(f)\cdot[\delta(f) + \sigma/2\delta(f-2f_{\textrm{ceo}}) +  \sigma/2\delta(f+2f_{\textrm{ceo}})],
\end{split}
\end{equation}
where $\mathcal{F}$ represents the Fourier transform operator, $P_0(f)$ is the unmodulated acoustic spectrum. Therefore, the sidebands appears at $\pm2f_{\textrm{ceo}}$. Figure 4(a) illustrates the CEP variation of our CEP-stabilized light pulses, corresponding to $f_\text{ceo} = 0$. Figures 4(b) and 4(c) show CEP variations with modulation frequencies of \(\approx 1.5\) Hz and \(\approx 2.5\) Hz, i.e., \( f_{\text{CEO}} \approx 1.5 \) Hz and \( f_{\text{CEO}} \approx 2.5 \) Hz, respectively. (Note: The y-axis ranges from \( \pi/2 \) to \( -\pi/2 \) instead of \( \pi \) to \( -\pi \), making the modulation appear as \(\approx 3\) Hz and \(\approx 5\) Hz.) Figure 4(d) presents the corresponding acoustic spectrum. Each acoustic harmonic exhibits identical \(\pm 2f_\text{ceo}\) sidebands. Here, we show only a zoomed-in view of the 42nd acoustic harmonic component. Clearly, when \( f_\text{ceo} \approx 1.5 \) Hz and \( f_\text{ceo} \approx 2.5 \) Hz, the sidebands shift to \( \pm2 \times 1.5 \) Hz \( \approx \pm3 \) Hz and \( \pm2 \times 2.5 \) Hz \( \approx \pm5 \) Hz, respectively. We also observe higher-order sidebands because of the acoustic amplitude's deviation from the assumed perfect cosine response to CEP changes. In short, acoustic sidebands appear at frequencies of \( n f_{\text{ceo}} \) (\( n = \pm~ 2, 4, 6, \dots \)), with the \( n = \pm2 \) sidebands being the most prominent. Note that the main acoustic harmonic peaks show barely any frequency shift, confirming that the CEP effect on waveform phase is minor, consistent with our time-domain observation in Figure 3(c). Despite the fact that this $f_{\text{CEO}}$ measurement method relies on air ionization (i.e., a few-cycle pulse with pulse energies of tens of microjoules), making it challenging for current MHz comb lasers, this new approach can be immediately applied in the strong-field and attosecond science community.

\begin{figure}[htbp]
\centering
\includegraphics[width=\linewidth]{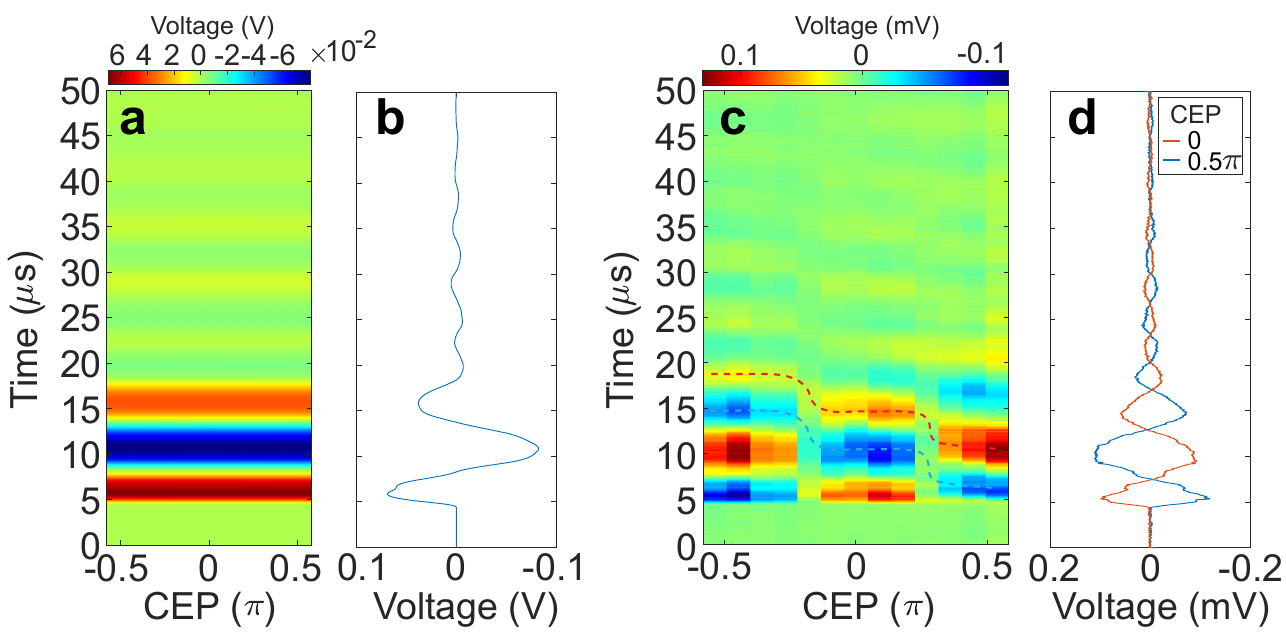}
\caption{\textbf{CEP dependence of acoustic waveform in the time domain.} CEP-resolved (\textbf{a}) and -averaged (\textbf{b}) acoustic waveforms. The colorbar in (a) represents the voltage. (\textbf{c}) CEP-resolved differential distribution by subtracting the CEP-averaged waveform. Two dashed curves have been added to highlight the stepwise $\pi$ phase shift. (\textbf{d}) Lineouts from (c) at the relative CEP = 0 and 0.5$\pi$.}
\label{fig:figure3}
\end{figure}

\begin{figure}[htbp]
\centering
\includegraphics[width=\linewidth]{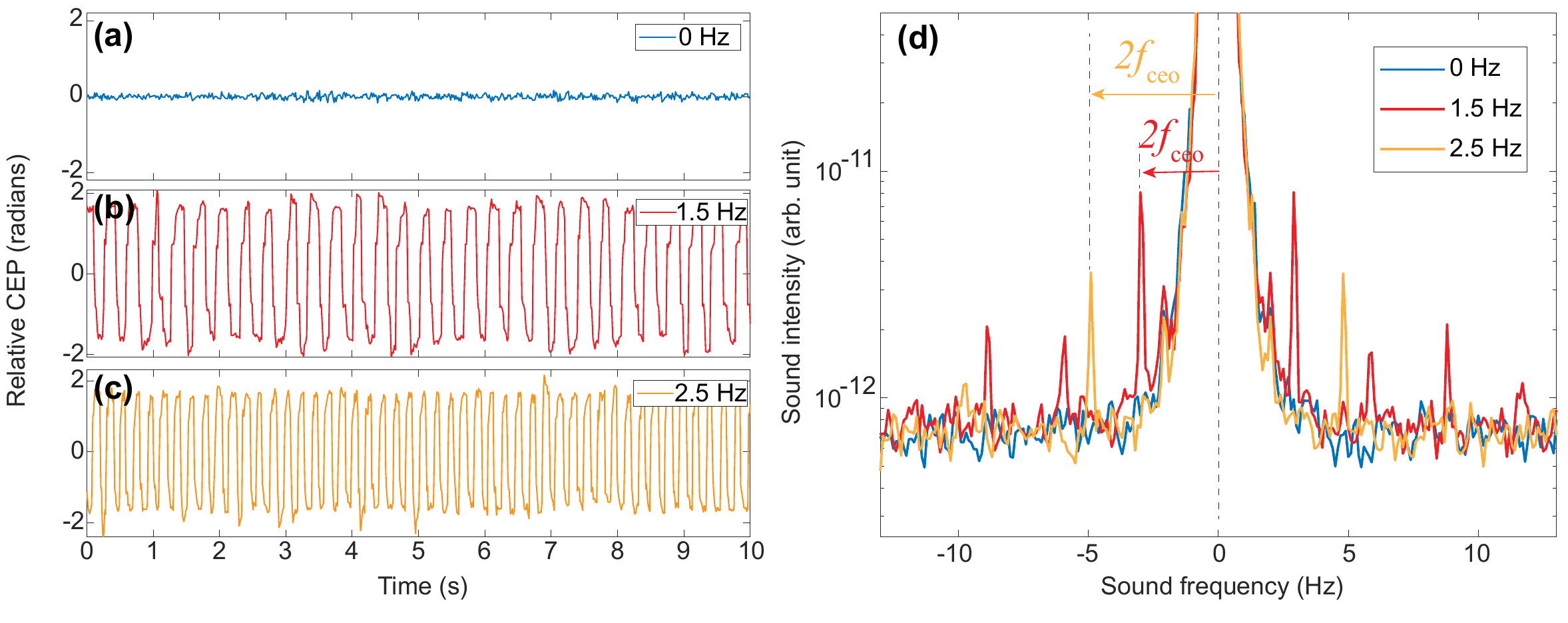}
\caption{\textbf{Hearing the carrier-envelope offset frequency.} (\textbf{a}) Stabilized and (\textbf{b}, \textbf{c}) modulated CEP as a function of time, with modulation frequencies set to 3 Hz and 5 Hz, respectively. Note that the modulation range is from $\pi/2$ to $-\pi/2$, and therefore the corresponding $f_{\text{ceo}}$ is 1.5 Hz and 2.5 Hz, respectively. (\textbf{d}) The corresponding acoustic spectrum. To observe the sidebands clearly, only a zoomed-in view of the 42nd acoustic harmonic is shown. Acoustic sidebands appear at \( n f_{\text{ceo}} \) (\( n = \pm~2, 4, 6, \dots \)). }
\label{fig:figure4}
\end{figure}

\section{Discussion and outlook}

The generation of the acoustic waves originates from the strong-field ionization. This is supported by the energy scaling rule of the acoustic intensity as a function of laser intensity presented in the supplemental document. A very similar trend with the ADK rate of tunneling ionization \cite{ammosov1986tunnel} was found. Therefore, the laser-acoustic wave generation can be understood through a two-step process \cite{kaleris2021laser}. The first step involves energy injection via strong-field ionization, which produces a large number of hot electrons and ions within the ionization volume. This process is common to both acoustic waves and fluorescence. The second step encompasses energy relaxation and temperature cooling. These relaxation processes occur on different time scales. Fluorescence typically involves energy relaxation on picosecond to nanosecond scales, as excited ions return to lower energy states by emitting photons. In contrast, acoustic waves are related to the macroscopic motion of the gas medium, such as expansion and compression, which occurs on a slower time scale.

Since the intensity of acoustic waves is sensitive to the electric field of a laser pulse—not only its intensity profile—we can use this property to characterize laser pulses. Recently, a straightforward waveform-characterization technique called TIPTOE (tunneling ionization with a perturbation for the time-domain observation of an electric field) has been demonstrated for measuring few-cycle pulses by collecting the ion current \cite{park2018direct}. Later, a variant by measuring the plasma fluorescence was demonstrated \cite{saito2018all,liu2021all,tsai2022nonlinear}. In our TIPTOE measurements, we used two identical pulses (except for the intensity) split from the same laser beam. The strong pulse ($\sim 1\times10^{14}$ W/cm$^2$) ionizes the air and thus gives rise to a laser plasma filament. A very weak pulse ($\sim 1\times10^{10}$ W/cm$^2$) perturbs the strong-field ionization, affecting the productions of the ionization, such as electron and ion yields, fluorescences and also the acoustic waves. Figure 5(a) shows the second-order acoustic spectrum as a function of time delay between the two pulses. Dominant oscillations around time zero correspond to the electric field of the main pulse, while oscillations beyond time zero are relate to the pedestal of the few-cycle pulse. Figure 5(b) shows the 391-nm fluorescence spectrum obtained from the TIPTOE measurement. These TIPTOE results consist of a cross-correlation DC background and a fast oscillation component. The latter reflects the temporal coherence of the laser beam. We use a period-averaged curve to represent the DC background and subtract it to extract the residuals, which correspond to the fast oscillation components. Figure 5(c) compares the residuals from the laser-acoustic wave and the 391-nm fluorescence. The close agreement between the two curves indicates that the laser-acoustic wave is a reliable observable for probing strong interactions in the laser-induced air plasma.

\begin{figure}[htbp]
\centering
\includegraphics[width=12cm]{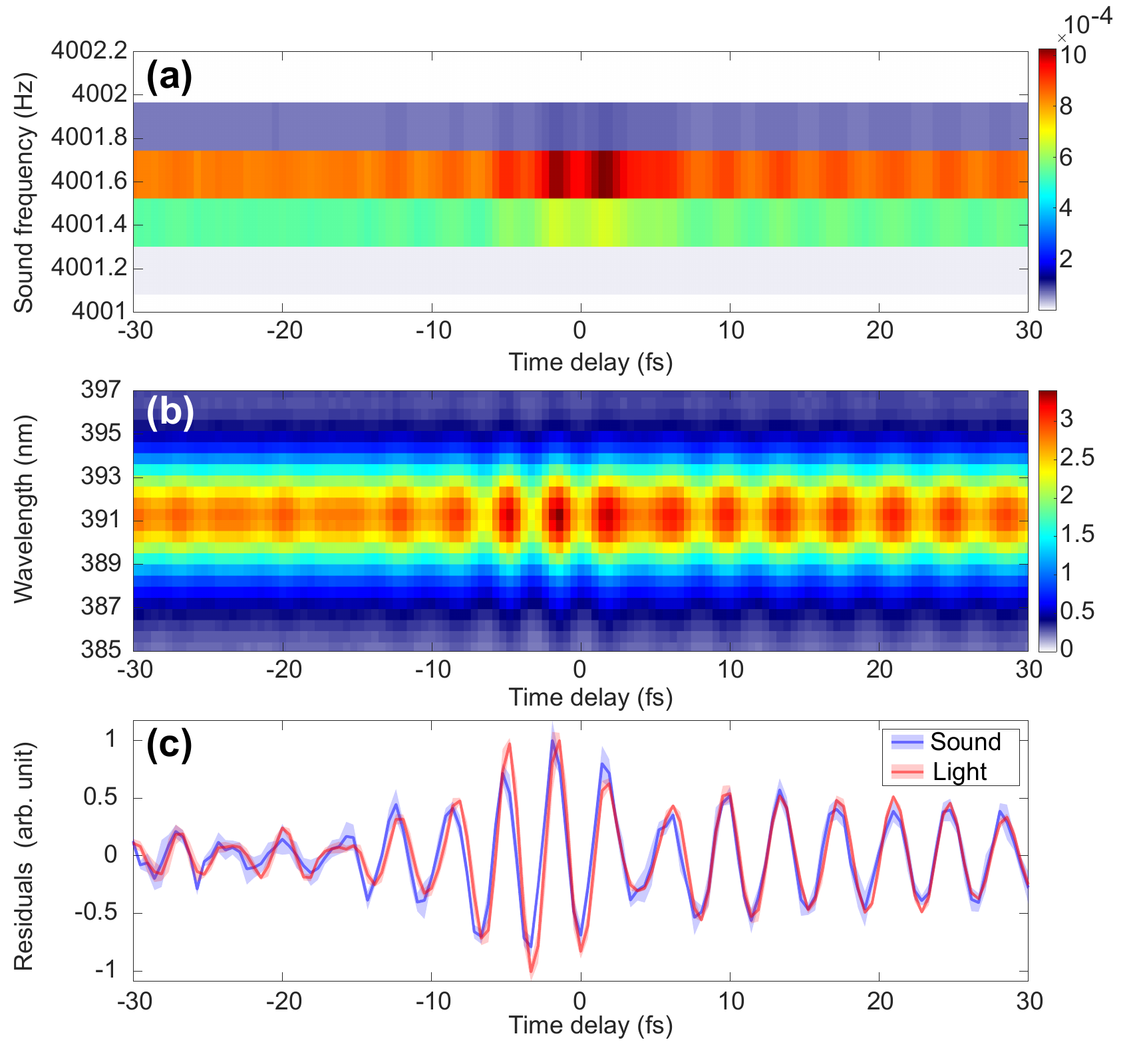}
\caption{\textbf{Pulse characterization based on laser acoustic waves.} (\textbf{a, b}) TIPTOE traces based on the 2nd acoustic spectral component and the 391-nm fluorescence, respectively. (\textbf{c}) Comparison of residual signals from (\textbf{a}) and (\textbf{b}), which are obtained by subtracting their dc backgrounds. The shade areas are the standard deviations of four measurements.}
\label{fig:figure5}
\end{figure}

%We believe our study only reveals the tip of the iceberg regarding laser-acoustic waves, yet it raises several open questions. For instance, how to numerically model the entire generation process, from \textcolor{blue}{ionized electron and ion bursts every half optical cycle} over collective plasma motions to millisecond gas pressure oscillations? Addressing this question will be crucial for advancing the spectroscopic applications based on these acoustic waves.

To conclude, we have demonstrated the first experimental observation of CEP-dependent acoustic waves induced by a few-cycle laser in air. Our results reveal that the acoustic waveform is modulated by the CEP, primarily affecting its amplitude, originating from laser-induced ionization. Leveraging this optoacoustic phenomenon, we introduce a novel method for directly sensing the carrier-envelope offset, $f_{\text{CEO}}$, frequency using a simple microphone. Furthermore, we also propose a new pulse characterization method—"hearing" the optical waveform. This optoacoustic phenomenon holds significant implications for frequency metrology and ultrafast science.

\begin{backmatter}
\bmsection{Funding}
This work was supported by the Chemical Sciences, Geosciences and Biosciences Division, Office of Basic Energy Sciences, Office of Science, US Department of Energy, grant no. DE-FG02-86ER13491. We thank the National Science and Technology Council, Taiwan, for funding grant no. 113-2112-M-007-042-MY3 to M.-C. C.

\bmsection{Acknowledgments}
We thank C. Aikens, S. Chainey and J. Millette for their technical support. We thank Dr. Huynh Lam for helping on the measurement with oscilloscope.

\bmsection{Disclosures}
M.H. disclosed a patent application.

\bmsection{Data availability}
Data underlying the results presented in this paper are not publicly available at this time but can be obtained from the authors upon reasonable request

\bmsection{Supplemental document}
See Supplement for supporting content. 

\end{backmatter}

%%%%%%%%%%%%%%%%%%%%%%% References %%%%%%%%%%%%%%%%%%%%%%%%%
%%%%%%%%%% If using BibTeX:
\bibliography{sample,references}

\end{document}